\documentclass[aps,prd,eqsecnum,11pt,showpacs,preprintnumbers,nofootinbib]{revtex4}

\usepackage{amssymb,amsmath}
\usepackage{graphicx}

\begin{document}

\title{Effects of quantized fields on the spacetime geometries of static spherically
 symmetric black holes}

\author{Paul R. Anderson}

\affiliation{Department of Physics, Wake Forest University, P.O. Box
7507, Winston-Salem, NC, 27109, U.S.A.
\;\; and\\
Racah Institute of Physics, Hebrew University of Jerusalem, Givat
Ram, Jerusalem, 91904, Israel \;\;
and \\
Departamento de F\'{\i}sica Te\'orica and IFIC, Universidad de
Valencia-CSIC, C. Dr. Moliner 50, Burjassot-46100, Valencia, Spain
}

\author{Mathew Binkley}
\affiliation{Department of Physics, Wake Forest University, P.O. Box
7507, Winston-Salem, NC, 27109, U.S.A.}

\author{Hector Calderon}
\affiliation{Department of Physics, Montana State University, Bozeman,
MT,  59717, U.S.A.}

\author{William A. Hiscock}
\affiliation{Department of Physics, Montana State University, Bozeman,
MT, 59717, U.S.A.}

\author{Emil Mottola}
\affiliation{Theoretical Division, T8, Los Alamos National Laboratory, Los Alamos,
NM, 87545, U.S.A.}

\author{Ruslan Vaulin}
\affiliation{Department of Physics, Florida Atlantic University, 777 Glades Road, Boca
Raton, FL, 33431, U.S.A.}

\begin{abstract}

Analytic approximations for the stress-energy of quantized
fields in the Hartle-Hawking state~\cite{H-H} in static black hole
spacetimes predict divergences on the event horizon of the black
hole for a number of important cases.  Such divergences, if real,
could substantially alter the spacetime geometry near the event
horizon, possibly preventing the black hole from existing.  The
results of three investigations of these types of effects are
presented.  The first involves a new analytic approximation for
conformally invariant fields in Reissner-Nordstr\"{o}m (RN)
spacetimes which is finite on the horizon. The second focuses on the
stress-energy of massless scalar fields in Schwarzschild-de Sitter
black holes. The third focuses on the stress-energy of massless
scalar fields in zero temperature black hole geometries that could
be solutions to the semiclassical backreaction equations near the
event horizon of the black hole.

\end{abstract}

\maketitle

It is well known that previous analytical approximations for
quantized massless fields in the Hartle-Hawking state in RN
spacetimes~\cite{F-Z,AHS} predict that one component of the
stress-energy tensor diverges logarithmically on the event horizon
in the nonextreme case.  For extreme Reissner-Nordstr\"{o}m (ERN)
black holes there is both a powerlaw and a logarithmic divergence
predicted.  On the other hand numerical calculations for massless
spin $0$ and spin $1/2$ fields~\cite{AHS,AHL,CHOAG} show no evidence
for such divergences.

For conformally invariant fields there is an approximate effective
action, sometimes called the anomaly action, which when varied with
respect to the metric results in a stress-energy tensor whose trace
is equal to the trace anomaly~\cite{Riegert,ft,os,mm}. The behavior
of this stress-energy tensor has been studied for Schwarzschild and
other spacetimes~\cite{bfs,mv}.  It can be written in terms of two
auxiliary fields which both obey fourth order differential
equations.  This results in numerous solutions which, at least to
some extent, correspond to various possible states for the quantized
fields~\cite{bfs,mv}.

An investigation of this action for the case of RN and ERN
spacetimes~\cite{AMV} shows that there exist solutions to the
auxiliary field equations which give a finite stress-energy on the
event horizons of both types of black holes.  The best approximation
can be obtained by fitting the value of ${T^t}_t$ on the horizon and
at large $r$ in RN.  This results in one free parameter which can be
varied to give the best fit to
  the stress-energy at intermediate locations.  The approximation
  for ERN is not as good because ${T^t}_t$ is fixed on horizon and the value is not close to the
  correct one.  However, it is still possible to
  fix the value of $({T^r}_r - {T^t}_t)/(-g_{tt})$ on the horizon and to fix ${T^t}_t$
  at large $r$.   While not perfect, these approximations have the
  advantage of being the only known approximations for massless quantized fields which do not
  have divergences on the horizon for RN and ERN.

The analytic approximation~\cite{AHS} for massless scalar fields
with arbitrary coupling $\xi$ to the scalar curvature $R$ makes the
somewhat surprising prediction~\cite{CHA} that if the black hole
temperature is nonzero and if $R \ne 0$ on the horizon (e.g. in
Schwarzschild-de Sitter spacetimes), then near the horizon
\begin{eqnarray} \langle {T^t}_t \rangle &\sim& - \langle {T^r}_r \rangle \sim \frac{ \xi (\xi -
1/6)}{r_h (r-r_h)} R(r_h)\;. \nonumber
\end{eqnarray}
These components are finite on the horizon in both RN and ERN
spacetimes where $R=0$. Given the fact that divergences predicted
for a different component of the stress-energy tensor in RN and ERN
turned out not to be real, one might expect that the same would be
true for these apparent divergences associated with $R \ne 0$.

To test this conjecture, numerical computations of the stress-energy
tensor for massless scalar fields with various values of $\xi$ have
been computed on the horizon for Schwarzschild-de Sitter
spacetimes~\cite{CHA}. In these spacetimes there is both an event
horizon for the black hole and a cosmological horizon and the two
are at different temperatures.  However, the fields can be put in
the Hartle-Hawking state if the black hole is surrounded by a
perfectly reflecting mirror.  The numerical calculations have been
done using a variation of a method~\cite{candelas} developed for
Schwarzschild spacetime that gives the values of $\langle {T^t}_t
\rangle$, $\langle {T^r}_r \rangle$, and $\langle {T^\theta}_\theta
\rangle$ on the horizon. The numerical evidence indicates that these
components are finite on the event horizon.

The existence of the trace anomaly, which is nonzero for
Schwarzschild and RN spacetimes, means that $R \ne 0$ for
self-consistent solutions to the semiclassical backreaction
equations.  The case of zero temperature black holes is a
particularly natural one to investigate because there is no mirror
needed to contain the radiation as there is for other black holes
when the fields are in the Hartle-Hawking state.

The existence of a divergence in the stress-energy of a potential
solution to the semiclassical equations would likely prevent that
solution from existing.  However, that does not necessarily mean
that no zero temperature black hole could exist.  Recent results
regarding zero temperature black holes in two dimensions
~\cite{bfffn,ff} show that it is possible that the divergences could
imply that no static state exists but that there is no divergence if
the black hole forms from gravitational collapse with the fields in
a particular nonstatic state.  Nevertheless, this is still a
significant deviation from one's intuition that static solutions to
the semiclassical backreaction equations should exist that describe
zero temperature black holes.

To determine the effects of the fields on the spacetime geometry in
as complete a way as possible, it is useful to work in the context
of a large $N$ expansion where $N$ is the number of identical
quantized scalar fields.  Unlike the usual loop expansion, the
leading order terms in the large $N$ expansion allow for the
possibility that quantum effects can significantly alter the
spacetime geometry.  It is also consistent to neglect the graviton
stress tensor since this comes in at next to leading order.

If only conformally invariant fields are present then it is possible
to argue that static spherically symmetric zero temperature black
hole metrics near the event horizon should have the general form
\[ ds^2 = -[a_2 (r-r_h)^2 + ...] dt^2 + [b_2 (r-rh)^2 + ...]^{-1} +
r ^2 d\Omega^2  \] with $r_h$ the radius of the event horizon and
$a_2$ and $b_2$ positive constants.
The trace equation has been solved in this case and it has been
found that $b_2$ is a function of $r_h$ and that $b_2
>1$ \cite{AM}.  There is good evidence that $\langle {T^\mu}_\nu \rangle$ depends
only on geometry near the horizon. Numerical results have been
obtained for the exact metric $ - g_{t t} = (r-r_h)^2/r^2$ and $g^{r
r} =  b_2 \, g_{tt}$ for various values of $b_2$.

In all cases considered so far, there is preliminary evidence that
the stress-energy of the conformally invariant scalar field diverges
on the horizon.
The divergence appears to occur for all values of $b_2
>1$.  However this is not conclusive since the value of the component which
diverges on the horizon depends on $a_3$, $b_3$ and possibly higher
terms in the expansion. In the cases where the divergence is present
it occurs for all $r_0
>0$ and thus for macroscopic as well as microscopic black holes.

In summary, a new analytic approximation has been found for
Reissner-Nordstr\"{o}m and extreme Reissner-Nordstr\"{o}m spacetimes
which is finite on the event horizon. No evidence has been found for
the existence of divergences predicted by an analytical
approximation for the stress-energy of quantized massless scalar
fields in nonzero temperature black hole spacetimes when $R \ne 0$.
There is evidence that spacetime geometries exist for which the
stress-energy of the conformally invariant scalar field diverges on
the event horizon of a zero temperature black hole.  However the
possibility that there are self-consistent static zero temperature
black hole solutions to the semiclassical backreaction equations has
not been ruled out.

\section*{Acknowledgments}
 P.R.A. would like to thank Jacob Bekenstein, Alessandro Fabbri,
 Sara Farese, and Jos\'e Navarro-Salas for helpful conversations.  This research has been partially
supported by grant numbers PHY-9800971, PHY-0070981, and PHY-0556292
from the National Science Foundation.  P.R.A. thanks the Gravitation
Group at the University of Maryland for hospitality and acknowledges
the Einstein Center at Hebrew University, the Forchheimer
Foundation, and the Spanish Ministerio de Educaci\'on y Ciencia for
financial support.


\end{document}